\documentclass[prl,aps,amsmath,amssymb,twocolumn,superscriptaddress]{revtex4}
\usepackage{times}
\usepackage{hyperref}
\usepackage{amsmath,amssymb}
\usepackage[usenames]{color}
\usepackage{amssymb}
\usepackage{grffile}
\usepackage[pdftex]{graphicx}
\usepackage{amsmath, amstext, amssymb, amsfonts, amsxtra}
\usepackage{textcomp}
\usepackage{xspace}
\DeclareSymbolFontAlphabet{\amsmathbb}{AMSb} 
\usepackage{soul}
\usepackage{xcolor}
\usepackage{epstopdf}
\usepackage{braket}
\usepackage{float}
\usepackage{tikz}
\usepackage{pgfplots}
\usepgfplotslibrary{statistics}
\usepackage{enumitem}
\setlist{nosep}
\usetikzlibrary{arrows}

\newcommand{\rhop}{\hat{\rho}} 
\newcommand{\sop}{\hat{\sigma}} 
\newcommand{\Vop}{\hat{V}}

\newcommand{\Di}{\mathcal{D}}

\newcommand{\Lop}{\mathcal{L}}
\newcommand{\IPR}{\mathcal{I}_{{\rm w}}} 
\newcommand{\AIPR}{\bar{\mathcal{I}}_{\rm w}}    
\newcommand{\SO}{\hat{T}} 
\newcommand{\tr}{{\rm tr}}

\newcommand{\Hop}{\hat{H}}

\newcommand{\im}{{\rm i}}

\newcommand{\su}{\uparrow} 
\newcommand{\sd}{\downarrow}

\newcommand{\sutd}{EPD Pillar, Singapore University of Technology and Design, 8 Somapah Road, 487372 Singapore}

\begin{document}

\title{Interplay of interaction and disorder in the steady state of an open quantum system}                     

\author{Xiansong Xu} 
\affiliation{\sutd}
\author{Chu Guo} 
\affiliation{\sutd}
\author{Dario Poletti}
\affiliation{\sutd}

\begin{abstract} 

Many types of dissipative processes can be found in nature or be engineered, and their interplay with a system can give rise to interesting phases of matter. Here we study the interplay among interaction, tunneling, and disorder in the steady state of a spin chain coupled to a tailored bath. We consider a dissipation which, in contrast to disorder, tends to generate a homogeneously polarized steady state. We find that the steady state can be highly sensitive even to weak disorder. We also establish that, in the presence of such dissipation, even in the absence of interaction, a finite amount of disorder is needed for localization. Last, we show that for strong disorder the system reveals signatures of localization both in the weakly and strongly interacting regimes.  
\end{abstract}

\maketitle

{\it Introduction.} The interplay of dissipation and interaction can give rise to rich physics from out-of-equilibrium phase transitions \cite{Mitra, DiehlZoller2008, DellaTorre, DiehlZoller2010} to complex relaxation dynamics \cite{PolettiKollath2012, CaiBarthel2012, PolettiKollath2013, SciollaKollath2015, GarrahanLesanovsky, MarcuzziLesanovsky2014, MarcuzziLesanovsky2015}.   
Of particular relevance, in the past few years, has been the use of tailored dissipation to engineer interesting states of matter \cite{DiehlZoller2008}. 
However, it is important to study the robustness and, more generally, the response of such states to disorder. 

It should be noted that the interplay of interaction and disorder (without dissipation) has also gathered vast interest. A disordered interacting system can show many-body localization (MBL) or be in a regime where the eigenstate thermalization hypothesis is valid \cite{BaskoAltshuler2006, OganesyanHuse2007, PalHuse2010}. MBL has been observed experimentally both in the presence of pure disorder or of a quasirandom potential \cite{DErricoModugno2014,Schreiber,Monroe,Bordia,Bordia2, Roushan2017} even in two dimensions \cite{Gross, BordiaBloch2017}. Disordered interacting systems have also been studied when in contact with a bath. 
Recent studies focused on a number of dissipative processes, both theoretically \cite{NandkishoreHuse2015, JohriBhatt2015, Nandkishore2015, FischerAltman2016, LeviGarrahan2016, EverestLevi2016, NandkishoreGopalakrishnan2016, HyattBauer2016, ZnidaricGoold2016, ZnidaricVarma2016, MedvedyevaZnidaric2016, DroennerCarmele2017, VanNieuwenburgFischer2017,KarlssonVerdozzi2018} and experimentally \cite{LuschenSchneider}, showing a rich phenomenology. 
However, the use of a suitable tailored bath which contrasts disorder and, in its absence, could generate peculiar quantum states has yet to be deeply investigated.

In the following, we concentrate on a spin-$\mathrm{1/2}$ chain, a common choice in the study of MBL. One important quantity which characterizes the effect of disorder is the difference in local magnetization between nearby spins. In fact, because of the presence of disorder, the local magnetization of a spin can be significantly different from that of its neighbors. 
We then consider the effects of a bath which, contrary to the effect of disorder, tends to reduce the difference in local magnetization between nearest-neighboring spins. Hence we expect a strong interplay between dissipation and disorder, which can be significantly influenced by the interaction. Such baths were first proposed in Ref. \cite{DiehlZoller2008} (although for bosons) and experimentally realized for spins in Ref. \cite{Schindler2013}. 
Recently this dissipation has been studied for a disordered quadratic bosonic chain \cite{Denisov, Denisov2} where it was shown that the steady state can be dominated by few localized modes.           

In this Rapid Communication, we first analyze the clean system, i.e., without disorder, and we show that its steady state has identically zero local magnetization on each site. We also prove analytically that the steady state for the case in which the interaction and the kinetic terms of the Hamiltonian are of the same strength is a highly symmetric entangled pure state which is sensitive to small amounts of disorder. 
We study the localization in the steady state by analyzing the natural orbitals of the single-particle reduced density matrix. 
This allows us to show a complex behavior of the indicators of localization with the strength of the interaction.

{\it Model.} We study the steady state of a dissipative spin chain described by a master equation in Lindblad form \cite{GoriniSudharsan, Lindblad}  
\begin{align}
    \label{eq:master}
    \frac{d\rhop}{dt}&=\Lop\left[\rhop\right]=-\frac{\im}{\hbar}\left[\Hop,\rhop\right]+\mathcal{D}\left[\rhop\right],  
\end{align}
where $\Lop$ is the system's Lindbladian. For the system's Hamiltonian we considered a Heisenberg $XXZ$ chain for spin-$\mathrm{1/2}$ with a spatially disordered local magnetic field $h_l$, 
\begin{align}
    \label{eq:ham}
    \Hop =\sum_{l=1}^{L-1}\left[J \left(\sop^{x}_l \sop^x_{l+1}+\sop^y_l \sop^y_{l+1}\right)+\Delta \sop^z_l \sop^z_{l+1}\right]+\sum_{l=1}^{L} h_l \sop^z_l, 
\end{align}
where the elements of $\sop_l^\alpha$ are given by the Pauli matrices for $\alpha=x,\;y$, or $z$ and where we have used open boundary conditions. The random field $h_l$ is uniformly distributed in $\left[-W,W\right]$ with $W$ characterizing the disorder strength. This model exactly maps to an interacting spinless fermionic chain under Jordan-Wigner transformation \cite{JordanWigner, Lieb}. The $XX$ part, parametrized by $J$, maps to the kinetic term for the fermionic chain, and the interaction $\Delta$ part maps to the nearest-neighbor interaction (we will thus refer to $\Delta$ as the interaction in the following). In the absence of dissipation, this model has been shown to have many-body localized or ergodic phases depending on the strengths of disorder and interaction \cite{PalHuse2010, BeraBardarson, Bera2017, LezamaBardarson, LinHeidrichMeisner2017, Alet, Abanin}. 
The dissipator we use is in Lindblad form and is given by       
\begin{align}
    \mathcal{D}\left[\rhop\right]&=\gamma\sum_{l=1}^{L-1} \left(\Vop_{l,l+1}\rhop \Vop_{l,l+1}^\dagger-\frac{1}{2}\left\lbrace \Vop_{l,l+1}^\dagger \Vop^{}_{l,l+1}, \rhop \right\rbrace \right),  \label{eq:diss}  
\end{align}
where $\gamma$ indicates the coupling strength, $\{.,.\}$ indicates the anticommutator, and the jump operators $\Vop^{}_{l,l+1}$ are     
\begin{align}
    \label{eq:jump} 
    \Vop^{}_{l,l+1}=\left(\sop^+_l+\sop^+_{l+1}\right)\left(\sop^-_l-\sop^-_{l+1}\right).
\end{align} 
The type of dissipator above has already been studied for bosonic and fermionic systems \cite{DiehlZoller2008,Yi, Bardyn}. Its possible realizations with ultracold atoms could rely, for example, on the immersion of the system in a superfluid bath and hence the interaction with Bogoliubov excitations  \cite{DiehlZoller2008}. Thanks to the use of a universal quantum computer made of ultracold ions, the dissipator we used in Eqs. (\ref{eq:diss}) and (\ref{eq:jump}) has been realized experimentally \cite{Schindler2013}. We note that such a realization, based upon a Trotterization of the evolution operator \cite{Lloyd1996}, is independent of the Hamiltonian evolution and it is thus completely unaffected by the presence of disorder. In Eq. (\ref{eq:jump}) we have used $\sop^+_l=\left(\sop^x_l+\im\sop^y_l\right)/2$ and $\sop^-=\left(\sop_l^x-\im\sop_l^y\right)/2$. From the expression of the jump operators we can observe that the dissipator favors a balanced magnetization on neighboring sites. 

To compute the steady-state $\rhop_s$ we either use exact diagonalization or evolve any initial state with a matrix product states'-based time evolution ($t$-MPS). Since both the Hamiltonian and the dissipator are number conserving, we use a number-conserving $t$-MPS algorithm which conserves, at the same time, the total magnetization both in the bra and in the ket portions of the density operator \cite{Lauchli,Bernier2017}. 
In this Rapid Communication we concentrated on the sector with $0$ total magnetization so as to probe strong effects of the interaction. 
The notation $\langle \hat{O} \rangle_i$ indicates ${\rm tr}(\hat{O}\rhop_s)$ for the $i$th disorder realization, whereas $\bar{O}$ is used for the average over $M$ disorder realization of the quantity $\langle \hat{O} \rangle_i$, i.e., $\bar{O}=\sum_{i=1}^M \langle \hat{O} \rangle_i/M$.

{\it Nondisordered case.} We first consider the system without disorder. It is easy to show that the steady state is invariant to an overall spin flip. 
Applying $\SO=\bigotimes_l \sop^x_l$ both to the right and to the left of $\Hop$ leaves it invariant $\SO\Hop\SO=\Hop$ for any value of $\Delta$ and $J$. Each jump operator $\Vop_{l,l+1}$ is instead turned into its opposite $\SO\Vop_{l,l+1}\SO=-\Vop_{l,l+1}$. If we now consider the steady state, we can write 
\begin{align}
0=\Lop(\rhop_s)=\SO\Lop(\rhop_{s})\SO=\Lop(\SO\rhop_{s}\SO). \label{eq:symm}
\end{align} 
Considering the $0$ total magnetization manifold and since the steady state is unique, Eq. (\ref{eq:symm}) implies that $\rhop_{s}=\SO\rhop_{s}\SO$. 
From this we can deduce that the local magnetization $\langle \sop^z_l \rangle=0$ at every site (note that since we are considering a disorderless case, we did not use the label $i$ for the average value). This is readily proven by 
\begin{align}
\langle \sop^z_l \rangle=\tr (\sop^z_l \rhop_{s}) = \tr \!\left[(\SO\sop^z_l\SO ) \; (\SO\rhop_{s}\SO) \right]= - \langle \sop^z_l \rangle, \label{eq:deltamagnetization}        
\end{align}
where we have used the fact that $\sop^x_l\sop^z_l\sop^x_l=-\sop^z_l$. In the presence of disorder, the Hamiltonian would not be invariant to the action of $\SO$, and hence it would be possible to have nonzero local magnetization. 
   
It is important to stress that the steady state becomes, for $\Delta=J$, a Dicke state, which is a highly symmetric entangled pure state. In fact the steady state is $\rhop_s=|\psi_S\rangle\langle\psi_S|$ where 
\begin{align}
|\psi_S\rangle=\frac{1}{\sqrt{R}}\sum_{\vec{r}}|\vec{r}\rangle.
\end{align}
Above $| \vec{r}\rangle=| r_1 r_2 \dots r_L \rangle $ with $r_l=0,1$, respectively, for $\vert\sd\rangle$ and $\vert\su\rangle$, denotes all the possible states at $0$ total magnetization, and $R$ is the total number of such combinations \cite{examplestates}. 
To prove that $|\psi_S\rangle\langle\psi_S|$ is the steady state for $\Delta=J$ we will make use of the fact that $\sop^+_l \vert r_l \rangle = (1-r_l)\vert 1-r_l \rangle$ and $\sop^-_j \vert r_l \rangle = r_l\vert 1-r_l \rangle $. For the dissipative part, it can be shown that each jump operator $\Vop_{l,l+1}$ acting on $|\psi_S\rangle$ gives $0$, in fact,       
\begin{align}
\Vop&_{l, l+1} \sum_{\vec{r}} \vert {\vec{r}} \rangle  \nonumber \\
=  &(\sop^+_l\sop^-_l - \sop^+_{l+1}\sop^-_{l+1} + \sop^-_l\sop^+_{l+1}-\sop^+_l\sop^-_{l+1})\sum_{\vec{r}} \vert {\vec{r}} \rangle \nonumber \\ 
= &\sum_{\vec{r}} (r_l - r_{l+1} - r_l(1-r_{l+1}) + (1-r_l)r_{l+1}) \vert {\vec{r}} \rangle = 0, 
\end{align} 
which proves that $\Di\left[|\psi_S\rangle\langle\psi_S|\right]=0$. For the Hamiltonian we have 
\begin{align}
&(\sop^x_l\sop^x_{l+1} + \sop^y_l\sop^y_{l+1} + \sop^z_l\sop^z_{l+1})\vert \psi_S \rangle = \vert \psi_S \rangle ,    
\end{align}
which implies that $\left[ \Hop, \vert \psi_S \rangle \langle \psi_S \vert \right] = 0 $ if $\Delta = J$. This highly symmetric pure state is also the ground state of an open boundary $XXX$ spin chain in the absence of dissipation if $J=\Delta<0$. Indeed this would not be the case in the presence of disorder. 

\begin{figure}
\includegraphics[width=.9\columnwidth]{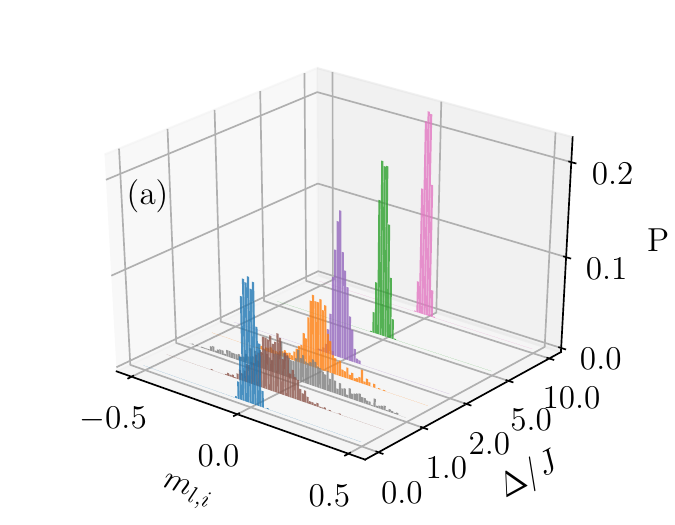}
\includegraphics[width=.9\columnwidth]{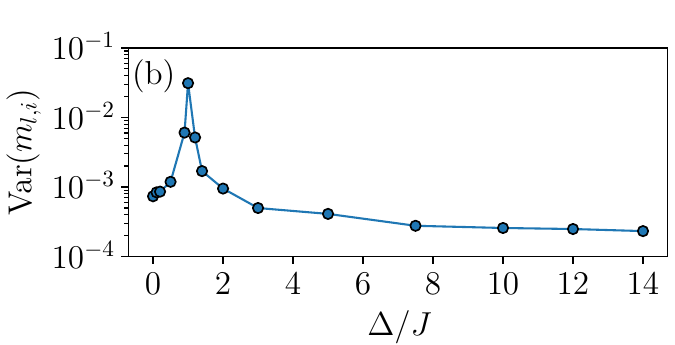}
\caption{(color online) (a) The distribution of local magnetization $P(m_{l,i})$ is plotted as a function of $m_{l,i}$ and of the interaction $\Delta$. The chains have lengths of $L=10$. The values of the interaction used are $\Delta/J=0,\;0.9,\;1,\;1.1,\;2,\;3,\;5$ , and $10$. The position of the distributions as a function of $\Delta$ is not linearly scaled for easier visualization. (b) Variance of the magnetization distribution $\mathrm{Var}(m_{l,i})$ as a function of the interaction $\Delta/J$. 
}  \label{fig:magnetization}             
\end{figure}

{\it Disordered case.} Disorder favors the occurrence of differences in the local spin orientation (magnetization), whereas the dissipator tends to reduce it. The local magnetization is defined as $m_{l,i}=\braket{\sop_l}_i$. We thus analyze the distribution of the local magnetization in the different sites for $100$ disorder realizations. A narrow distribution indicates a small difference in the local magnetization for all sites. If instead the local magnetization varies significantly, a broad distribution would be expected. In the absence of disorder, as we have shown in the previous paragraphs, the local magnetization is zero for all sites. 

We first consider a small magnitude of disorder $W=0.1J$, and we study its interplay with the interaction. The magnetization distribution profiles are represented in Fig. \ref{fig:magnetization}(a). Given the small amount of disorder, we notice that for $\Delta=0$ the spread of magnetization is fairly narrow. We then observe that the distribution becomes broader as the interaction increases (hence enhancing the role of disorder), whereas for larger interactions the distribution becomes narrow again. The magnetization distribution at large interaction becomes particularly narrow as the interaction dominates over kinetic energy and disorder. The change in the local magnetization distribution is more quantitatively characterized by the variance of the distributions, which is shown in Fig. \ref{fig:magnetization}(b). Here a peak of the variance is evident at $\Delta \approx J$. 
The effect of disorder is thus stronger for $\Delta \approx J$ where the disorderless Hamiltonian and the dissipator tend to produce the steady-state $\rhop_s=|\psi_S\rangle\langle\psi_S|$ showing that this state is much less robust against disorder.

\begin{figure}
\includegraphics[width=.9\columnwidth]{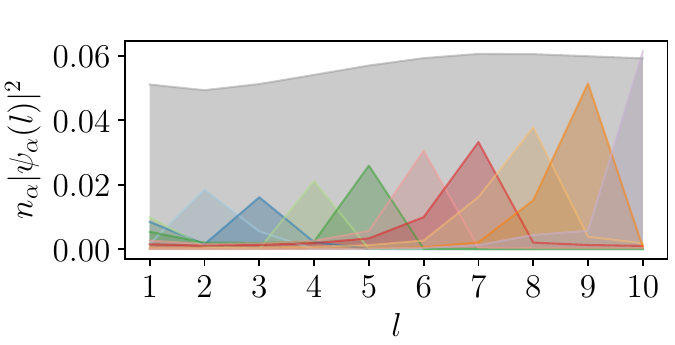}
\caption{(color online) Weighted natural orbitals as a function of position $l$ for a single disorder realization. Parameters used are $L=10$, $\Delta=J$, and $\hbar\gamma=W=0.1J$.}
\label{fig:nonlocal}
\end{figure}

To obtain a further understanding on the broadening of local magnetization for $\Delta \approx J$, we study the single-particle correlations to build the single-particle reduced density matrix $\rho_{\mathrm{sp}}^{j,k}=\langle \sop^+_j \sop^-_k\rangle$  \cite{BeraBardarson, Bera2017, LezamaBardarson, LinHeidrichMeisner2017}. Note that here and in the following, to lighten the notations, we have dropped subindex $i$ to indicate the disorder realization. From $\rho_{\mathrm{sp}}$ we compute its natural orbitals $\psi_{\alpha}$, i.e., the eigenvectors such that $\rho_{\mathrm{sp}}\psi_{\alpha}=n_{\alpha}\psi_{\alpha}$ where $n_{\alpha}$ is the normalized occupation spectrum. The natural orbitals in this regime are shown in Fig. \ref{fig:nonlocal} where a drastic contrast is shown among a fully delocalized orbital and all the other orbitals with large single-site occupation \cite{nonexponential}. This representation of the natural orbital is indicative and consistent with the presence of different local magnetizations.

\begin{figure}
\includegraphics[width=.9\columnwidth]{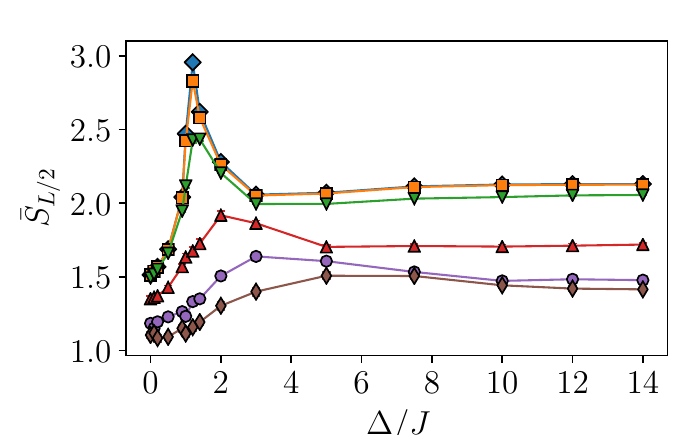}
\caption{Averaged operator space entanglement entropy computed for the middle of the chain $\bar{S}_{L/2}$ versus interaction $\Delta$ for different disorder strengths from top to bottom $W/J=0$ (the blue diamond), $0.1$ (the orange square), $0.5$ (the green down triangle), $2$ (the red up triangle), $5$ (the purple circle), $10$ (the brown thin diamond). The other parameters are $L=10$ and $\hbar\gamma=0.1J$. The error bars are comparable to the marker size. 
}  \label{fig:osee} 
\end{figure}

The strong effect of disorder for $\Delta\approx J$ can also be observed by studying the operator space entanglement entropy (OSEE) \cite{OSEE}, which we refer to as $\bar{S}_l$. This is a generalization to open systems of the bipartite entanglement entropy. From a matrix product operator representation of the steady state \cite{Schollwock2011, VerstraeteCirac2004, ZwolakVidal2004}, the OSEE is computed from the singular values of a bipartition of the system at site $l$, $s_{\alpha,l}$ in the following way 
\begin{align}
S_l=-\sum_{\alpha} \frac{s^2_{\alpha,l}}{\sum_{\alpha}s^2_{\alpha,l}}\ln\left( \frac{s^2_{\alpha,l}}{\sum_{\alpha}s^2_{\alpha,l}} \right). \label{eq:OSEE}
\end{align}
In Fig. \ref{fig:osee} we show $\bar{S}_{L/2}$ for the middle of a chain with $L=10$ as a function of the interaction and for different disorder strengths (lower to larger from top to bottom) \cite{n_realizations}. For the nondisordered case the OSEE is maximum for $\Delta=J$, and then $\bar{S}_{L/2}$ decreases significantly and increases again with the interaction.    
Disorder strongly suppresses the peak in $\bar{S}_{L/2}$ before lowering when it is strong also the other parts of the curve (see the bottom lines of Fig. \ref{fig:osee}).


\begin{figure}
\includegraphics[width=.9\columnwidth]{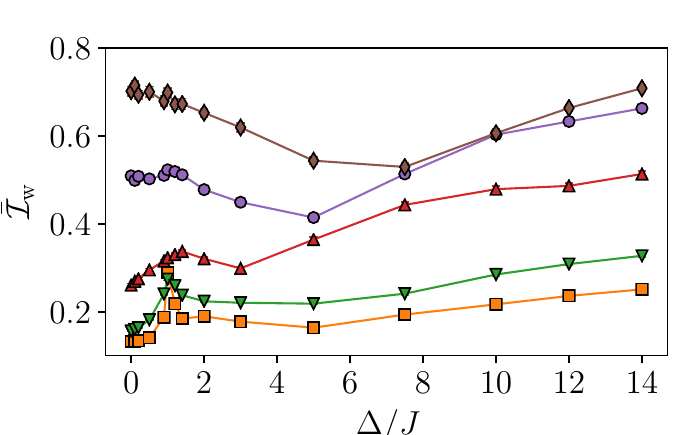}
\caption{(color online) Weighted averaged inverse participation ratio $\AIPR$ as a function of interaction for different disorder strengths: from bottom to top $W/J=0.1$ (the orange squares), $0.5$ (the green down triangles), $2$ (the red up triangles), $5$ (the purple circles), $10$ (the brown thin diamond). The other parameters are $L=10$ and $\hbar\gamma=0.1J$. The error bars are comparable to the marker size.} \label{fig:IPR} 
\end{figure}
\begin{figure}
\includegraphics[width=.9\columnwidth]{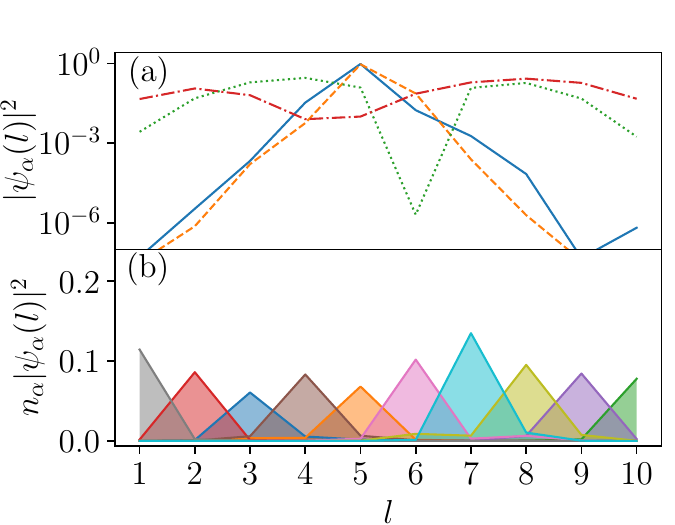}
\caption{(color online) (a) Single natural orbitals $|\psi_{\alpha}(l)|^2$ in logarithmic scale as a function of position $l$ for $W=5J$, $\Delta=14J$ (the blue solid line), $W=10J$, $\Delta=0.1J$ (the orange dashed line), $W=0.1J$, $\Delta=10J$ (the green dotted line), $W=0.1J$, $\Delta=0.1J$ (the red dot-dashed line) for a single realization. (b) Weighted natural orbitals $n_\alpha|\psi_{\alpha}(l)|^2$ as a function of position $l$ for a single disorder realization. The parameters used are $L=10$, $W=5J$, $\Delta=14J$, and $\hbar\gamma=0.1J$.   } \label{fig:localization} 
\end{figure}

A natural question to address is that of whether the steady state of the system has signatures of localization and how they are affected by the interaction. Again, we study the single-particle reduced density matrix $\rho_{\mathrm{sp}}^{j,k}=\langle \sop^+_j \sop^-_k\rangle$  and its orbitals $\psi_{\alpha}$. 
We then compute a weighted inverse participation ratio $\IPR=\sum_{\alpha,l} n_{\alpha}|\psi_{\alpha}(l)|^4$ and its average $\AIPR$ over the disorder realizations. If $\AIPR$ is close to $1$, it means that the relevant natural orbitals are localized, whereas for lower values of $\AIPR$, some relevant natural orbitals are localized \cite{nonweighted}. In Fig. \ref{fig:IPR} we depict $\AIPR$ versus the interaction $\Delta$ for different disorder magnitude strengths $W$, increasing from the bottom to the top curve. 
The dependence of $\AIPR$ with $\Delta$ is nonmonotonous and it is a function of the strength of disorder. At weak disorder, small interaction ``favors'' disorder increasing $\AIPR$ until, at intermediate values of the interaction, $\AIPR$ decreases as $\Delta$ increases. 
For even larger $\Delta$, $\AIPR$ rises, and for large enough interaction and disorder, all the natural orbitals of the steady state are localized, each with a different weight $n_{\alpha}$ \cite{fermiliquid}.            
To clearly illustrate this, in Fig. \ref{fig:localization}(a) we plot a typical relevant natural orbital as a function of position for four cases: weak disorder and weak interactions (the red dot-dashed line), weak disorder and strong interactions (the green dotted line), strong disorder and weak interactions (the orange dashed line) and intermediate strength of disorder and strong interactions (the blue solid line) \cite{alpha}. The first two cases (weak disorder) are delocalized, whereas the other two are localized. We deduce that intermediate to strong disorder is needed for localization as both cases with weak disorder have delocalized orbitals. Note also that for weak disorder $\AIPR$ in Fig. \ref{fig:IPR} is small for all values of the interaction $\Delta$. For intermediate or strong disorder, the natural orbitals can be localized, in particular, also in the presence of strong interaction. We stress that, for intermediate disorder $W = 5J$, strong interactions, e.g., $\Delta = 14J$, favor localization of the orbitals. It is however also important to consider the occupation of the orbitals. The interplay of disorder, dissipation, and interaction is such that, when the effects of disorder are strong enough, each natural orbital is exponentially localized and has a different occupation. This disordered scenario with localized orbitals is shown in Fig. \ref{fig:localization}(b) in which all the natural orbitals $\psi_{\alpha}$ are multiplied by their weight $n_{\alpha}$ for a typical disorder realization.

{\it Conclusions.} We have considered an interacting spin chain to study the interplay among tailored dissipation, interaction, tunneling, and disorder. The dissipation is such that it contrasts disorder and, in fact, we have proven that without disorder the steady state has zero magnetization on each site, independent of the strength of the interaction. Moreover, in the limit in which our spin chains reverts to a $XXX$ chain, the steady state is an entangled highly symmetric pure state. 
In this regime we have shown that the steady state can be very sensitive even to small amounts of disorder.
In the presence of strong disorder, the steady state has localization signatures indicated by localization of all the natural orbitals of the single-particle reduced
density matrix, a large inverse participation ratio, and a different occupation for each orbital. In this case, small and intermediate interactions can lower the inverse participation ratio, but large interactions can enhance it, indicating that all natural orbitals can be localized even at large interactions.  

More work is required to better understand dissipative disordered many-body systems, the phases that emerge and their properties, especially, in the spirit of dissipative engineering, with the use of tailored baths. The role of external time-dependent drivings to probe or further enrich these systems could be another interesting research direction.

Concurrently with our Rapid Communication, another study of a similar model whose results are consistent with ours appeared on the arXiv \cite{VakulchykDenisov2017}.       

{\it Acknowledgments.} D.P. acknowledges support from the Ministry of Education of Singapore AcRF MOE Tier-II (Project No. MOE2016-T2-1-065) and fruitful discussions with S. Denisov and F. Heidrich-Meisner. D.P. was hosted by ICTP (Trieste) and PCS-IBS (Daejeon) during part of this study. The computational work for this Rapid Communication was performed on resources of the National Supercomputing Centre, Singapore \cite{nscc}.


\begin{thebibliography}{10}
\bibitem{examplestates} For four spins, $R=6$ and the possible states are $|\sd\sd\su\su\rangle$, $|\sd\su\sd\su\rangle$, $|\sd\su\su\sd\rangle$, $|\su\sd\sd\su\rangle$, $|\su\sd\su\sd\rangle$ and $|\su\su\sd\sd\rangle$.
\bibitem{nonexponential} Here, we stress that these seemingly localized orbitals with large single-site occupation are not exponentially localized.
\bibitem{nonweighted} We have also computed the nonweighted inverse participation ratio (substituting $n_\alpha$ with $1/L$ in the expression for $\IPR$) which is qualitatively similar to the weighted one, showing significant quantitative differences only for low disorder.
\bibitem{alpha} The natural orbitals for the weak disorder cases are randomly chosen. For the strong disorder cases, the two natural orbitals are chosen such that they are located in the middle of the chain so as to show both exponentially decaying tails.
\end{thebibliography}
\end{document}